\documentclass{article}

\usepackage{arxiv}
\usepackage[numbers,sort&compress]{natbib}
\usepackage{tabularray}
\usepackage{adjustbox}
 \usepackage{amsmath}
\usepackage[utf8]{inputenc} 
\usepackage[T1]{fontenc}    
\usepackage{hyperref}       
\usepackage{url}            
\usepackage{booktabs}       
\usepackage{amsfonts}       
\usepackage{nicefrac}       
\usepackage{microtype}      
\usepackage{lipsum}		
\usepackage{graphicx}
\usepackage{natbib}
\usepackage{tabularray}
\usepackage{adjustbox}
\usepackage{amsmath}
\usepackage{graphicx}
\usepackage{placeins}
\usepackage{doi}
\usepackage{hyperref}
\usepackage{booktabs, multirow}

\title{EyeMakeYou: Identity-, Task-, and Subjective-State-Conditioned Diffusion for
High-Frequency Gaze Synthesis}
\renewcommand{\headeright}{}
\renewcommand{\undertitle}{}
\renewcommand{\shorttitle}{}

\date{} 					

\author{
Kamrul Hasan, Mehedi Hasan Raju, Oleg V. Komogortsev\\
Texas State University, San Marcos, Texas, USA\\
{\tt\small \{kamrul.hasan, m.raju, ok\}@txstate.edu}
}

\renewcommand{\shorttitle}{EyeMakeYou: Identity-, Task-, and Subjective-State-Conditioned Diffusion for
High-Frequency Gaze Synthesis}

\hypersetup{
pdftitle={A template for the arxiv style},
pdfsubject={q-bio.NC, q-bio.QM},
pdfauthor={David S.~Hippocampus, Elias D.~Striatum},
pdfkeywords={First keyword, Second keyword, More},
}

\begin{document}
\maketitle

\begin{abstract}
Eye movement biometrics (EMB) is an emerging behavioral modality for user authentication, particularly in virtual- and augmented-reality systems, where gaze dynamics contain distinctive subject-specific features. However, robust EMB systems require diverse, high-quality gaze recordings that are expensive to collect and often unavailable at the scale needed for model development. Generative models can mitigate data scarcity, but existing methods either synthesize generic gaze behavior or personalize signals primarily by identity, without jointly representing the user's task and subjective state. Consequently, generated signals may appear visually realistic while failing to retain the behavioral properties required for biometric applications. To address this limitation, we propose EyeMakeYou, a multi-conditional denoising diffusion framework for subject-specific, high-frequency gaze synthesis. EyeMakeYou generates 5-s, 1000-Hz bivariate gaze-velocity sequences from an identity-removed reference trajectory and conditions the denoising process on an identity embedding, a task embedding, and self-reported ratings of overall difficulty, mental tiredness, and eye tiredness. Its objective combines diffusion noise prediction and identity preservation with multi-resolution spectral, drift-consistency, and event-weighted local-smoothness losses. Experiments on GazeBase show that EyeMakeYou achieves higher median spatial accuracy and greater real--synthetic similarity in the embedding feature space than the existing generative approaches, while retaining selected task-dependent associations between subjective reports and oculomotor features. These findings support conditional diffusion as a practical approach for augmenting gaze datasets for biometric and interactive applications.
\end{abstract}


\section{Introduction}
Biometrics refers to measurable physiological and behavioral characteristics used to recognize individuals \cite{jain2004introduction}. Most existing biometric modalities for human recognition are commonly classified into physiological and behavioral categories. Physiological biometrics relies on an individual's physical traits, including facial features, fingerprints, iris patterns, palmprints, and retinal patterns \cite{alsaadi2015physiological}. These traits are generally stable over time, although the degree of user cooperation required to capture them depends on the sensing setup and application. In contrast, behavioral biometrics reflects characteristic patterns of human activity, with common modalities including gait, signature, keystroke dynamics, and voice; because such signals can often be captured without 'subjects' cooperation, they are well suited to continuous authentication \cite{alzubaidi2016authentication}.

Eye movement biometrics (EMB) is a relatively recent behavioral biometric modality that can support user authentication through person-specific characteristics of the complex oculomotor system. Beyond gaze-based authentication, eye-movement analysis has also been investigated for liveness detection and for identifying gaze patterns associated with dyslexia and autism spectrum disorder (ASD) \cite{thanarajan2023eye}. In addition to these applications, eye tracking facilitates foveated rendering \cite{patney2016towards} in augmented and virtual reality (AR/VR), enhances adaptive interfaces \cite{menges2019improving}, and supports attention-aware tutoring \cite{hutt2021breaking}. In transportation and safety, gaze and pupil metrics can support monitoring of fatigue and distraction \cite{singh2011eye}, while in behavioral science and marketing, they can reveal patterns of attention, decision-making, and expertise \cite{fiedler2012dynamics}. These applications illustrate the broad value of eye movements as a behavioral signal and of eye tracking as a versatile tool for clinical research and robust privacy-aware human-computer interaction (HCI).

However, applications that depend on fine oculomotor dynamics require high-sampling-rate eye-movement signals (e.g., 250--1000~Hz) to represent rapid events and subtle temporal structure, including saccades, microsaccades, and fine-grained smooth pursuit. These signals also contain subject-specific information that may enable re-identification and inference of sensitive attributes, including emotional state, gender, and ethnicity, creating privacy and ethical concerns \cite{khan2019survey, steil2019privacy}. Furthermore, collecting high-quality, high-frequency gaze data is resource-intensive, requiring specialized equipment, participant time, and labor \cite{griffith2021gazebase}. Moreover, many large-scale datasets remain proprietary or not available for public use \cite{garbin2020dataset}. This motivates subject-specific synthetic gaze generation as a solution to augment limited real datasets \cite{hasan2026diffusion, prasse2023sp}.

Synthetic gaze refers to artificially generated gaze signals intended to preserve the properties of real gaze that are relevant to a target application. Recently, a range of deep generative models, including generative adversarial networks (GANs), variational autoencoders (VAEs), and denoising diffusion probabilistic models (DDPMs), have been used to synthesize eye-movement signals. For example, SP-EyeGAN is a GAN-based method that generates sequences containing fixation events (periods of relatively stable gaze) and saccade events (rapid movements between fixation locations) \cite{prasse2023sp}. In SP-EyeGAN, synthesis starts from a random noise distribution rather than an explicit subject-specific condition. DiffEyeSyn, in contrast, uses a conditional diffusion model that takes an identity-removed velocity reference and a user embedding to generate user-specific 5-s signals \cite{jiao2024diffeyesyn}. More recently, Hasan et al. provided a quantitative comparison of GAN- and diffusion-based approaches for gaze synthesis \cite{hasan2025quantitative}. However, these approaches do not jointly condition generation on user identity, task context, and subjective state; consequently, they cannot separately specify all factors that may influence gaze behavior.

To overcome these limitations, we propose EyeMakeYou, a multi-conditional diffusion model for synthesizing subject-specific, high-frequency gaze-velocity signals. The model incorporates three complementary conditions: user identity, task context, and subjective user state. First, we construct a low-frequency conditional reference by downsampling each real gaze-position signal to 20~Hz and linearly interpolating it back to the original sampling rate of 1000~Hz. This operation retains the coarse trajectory while attenuating higher-frequency components that may contain identity-related features \cite{raju2021determining}. Second, we use pre-trained Eye Know You Too (EKYT) encoders to extract a 128-dimensional identity representation from the original velocity signal. Later, the task context is represented by a learned 128-dimensional embedding associated with the task performed, such as text reading or random saccades. We also incorporate three self-reported subjective-state ratings---overall difficulty, mental tiredness, and eye tiredness---on a 1--7 Likert scale, where 1 denotes the lowest and 7 the highest reported level. These ratings are encoded by a two-layer multilayer perceptron to produce a 32-dimensional state representation. To improve signal fidelity beyond the standard diffusion noise-prediction objective, we introduce complementary training losses that operate at multiple levels. Specifically, the multi-resolution spectral loss matches frequency-domain structure across temporal scales, the drift-consistency loss reduces accumulated displacement error after velocity integration, and the event-weighted local-smoothness loss matches local temporal curvature during fixation and post-saccadic oscillations. Together, the low-frequency reference and identity-, task-, and state-related conditions provide the denoiser with explicit information for generating more authentic, subject-specific gaze signals.

Finally, we conduct a rigorous evaluation of synthetic gaze using complementary spatial, biometric, and state-related measures. Signal quality is assessed through spatial accuracy and spatial precision, whereas biometric preservation is assessed using cosine similarity between real and synthetic EKYT embeddings. We further evaluate state-related fidelity by examining associations between subjective-state reports and oculomotor features. To make the spatial findings actionable for system design, we adopt the user-centric evaluation strategy of \cite{aziz2024evaluation}, which summarizes performance across user percentiles ($U$) and within-user error percentiles ($E$). This framework yields interpretable operating points: $U50|E50$ represents the median user under median within-user error, whereas $U95|E95$ represents the high-error operating point associated with the 95th percentile of users and within-user samples.

The major contributions of our study are summarized as follows:

\begin{enumerate}

\item We propose an identity-, task-, and user-state-conditioned diffusion model for generating subject-specific synthetic gaze signals.

\item We jointly optimize noise-prediction, identity-preservation, multi-resolution spectral, drift-consistency, and event-weighted local-smoothness losses to generate gaze sequences that are biometrically distinctive, spectrally faithful, drift-stable, and physiologically plausible.



\end{enumerate}

\section{Related Work}
\subsection{Statistical Models}
Early approaches to eye-movement synthesis typically used training-free statistical, signal-processing, or computer-graphics models \cite{lee2002eyes, duchowski2015modeling, duchowski2016eye}. For example, Lee et al. \cite{lee2002eyes} proposed an eye-movement model for animated faces based on observations of human saccades and statistical patterns in eye-tracking data. Later, Ma et al. \cite{ma2009natural} modeled the coordination between head motion and gaze direction to generate natural eye movements from a given head trajectory. Le et al. \cite{le2012live} extended multimodal synthesis by mapping speech features to simultaneous eye, head, and eyelid movements in a high-dimensional feature space. Addressing a related problem, Wood et al. \cite{wood2015rendering} introduced SynthesEyes, which renders photorealistic eye images with corresponding gaze labels to support the training of eye-tracking models. Together, these studies established useful foundations for modeling and rendering gaze-related behavior. Still, they generally did not target subject-specific, high-frequency oculomotor dynamics such as drift, tremor, and microsaccades.

Subsequent statistical approaches moved closer to raw eye-movement synthesis by incorporating explicit oculomotor characteristics \cite{le2012live, fuhl2018eye_Kasneci, fuhl2018eye_Santini, lan2022eyesyn}. Yeo et al. \cite{yeo2012eyecatch} introduced EyeCatch, which uses a Kalman filter \cite{welch1995introduction} to simulate sequences of fixations, saccades, and smooth pursuits. Campbell et al. \cite{campbell2014saliency} proposed a unified Bayesian model to estimate scanpath dynamics and saliency jointly, identifying distinct saliency and gaze patterns between children with autism spectrum disorder (ASD) and typically developing (TD) children. More recently, Lan et al. \cite{lan2022eyesyn} proposed EyeSyn, a physics-based synthesis model that generates fixations, saccades, and smooth pursuits using psychology-inspired equations and introduces ocular jitter through Gaussian drift and pink noise. Although these methods incorporate explicit oculomotor features, they typically produce generic behavior and may not capture individual differences in eye-movement dynamics. They also often target scanpath-level behavior or represent fine high-frequency components, such as microsaccades and ocular tremor, through simplified assumptions or injected noise. Consequently, statistical models may be insufficient when realistic, high-resolution, individualized gaze signals are required, motivating data-driven machine-learning approaches that can learn complex eye-movement dynamics from data.

\subsection{Machine-learning Models}
To address the limitations of statistical gaze-synthesis models, researchers have increasingly explored data-driven machine-learning approaches, including convolutional, adversarial, variational, and diffusion-based models \cite{simon2016automatic, assens2018pathgan, goodfellow2020generative, kaur2020eyegan, jiao2023supreyes}. Early image-conditioned work combined visual input with sequence modeling to predict human-like scanpaths. For example, Simon et al. \cite{simon2016automatic} employed a CNN--LSTM architecture that takes a static image and predicts an eye-movement sequence consisting of fixations and saccades, thereby modeling typical visual scanning behavior. Similarly, Assens et al. \cite{assens2018pathgan} proposed PathGAN, which predicts likely fixation locations from an image. Although PathGAN models the spatial allocation of attention, it produces discrete fixation positions rather than continuous gaze trajectories and therefore does not represent saccadic trajectories or velocity profiles. Beyond image-conditioned models, several studies \cite{fuhl2021fully, de2022next, fuhl2022hpcgen} investigated stimulus-agnostic approaches for synthesizing gaze data. Fuhl et al. \cite{fuhl2021fully} used fully convolutional networks for semantic segmentation, reconstruction, and variational autoencoder (VAE)-based generation of raw eye-tracking data without preprocessing, thereby allowing flexible input dimensions. Because this approach is not explicitly conditioned on user identity or task context, it generates generic rather than subject-specific behavior. More recent work has applied generative adversarial networks \cite{goodfellow2020generative} and diffusion models \cite{ho2020denoising, nichol2021improved} to eye-movement synthesis. SP-EyeGAN \cite{prasse2023sp} is a GAN-based method that explicitly targets fixation and saccade events, but it generates sequences from a general distribution of reading-related eye movements rather than conditioning the output on a specific user. Jiao et al. \cite{jiao2025diffgaze} introduced DiffGaze, a diffusion-based model that synthesizes 30-Hz scanpaths for 360\textdegree{} virtual-reality images. Although DiffGaze provides stimulus-driven gaze data for virtual environments, it does not explicitly model subject identity.

Meanwhile, \cite{jiao2024diffeyesyn} introduced a conditional diffusion framework for user-specific, high-frequency gaze synthesis. Built on DiffWave \cite{kong2020diffwave} and conditional diffusion \cite{zhang2023adding}, the model uses a downsampled-and-interpolated identity-suppressed velocity reference together with an EKYT-derived 512-dimensional user embedding \cite{lohr2022eye} to generate user-specific signals. This design is a crucial step toward personalized gaze synthesis; however, high-dimensional identity embeddings increase the complexity of the conditioning input. Later, Hasan et al. present a comparative study of GAN- and diffusion-based subject-specific gaze synthesis and their effectiveness for biometric utility \cite{hasan2025quantitative}. In the present work, we build on personalized diffusion-based gaze synthesis by jointly conditioning the denoiser on identity, task, and subjective user-state representations. This formulation makes the intended user, task context, and reported state explicitly available during generation, enabling a more comprehensive treatment of subject-specific gaze synthesis.

\begin{figure*}[ht]
    \centering
\includegraphics[width=\linewidth]{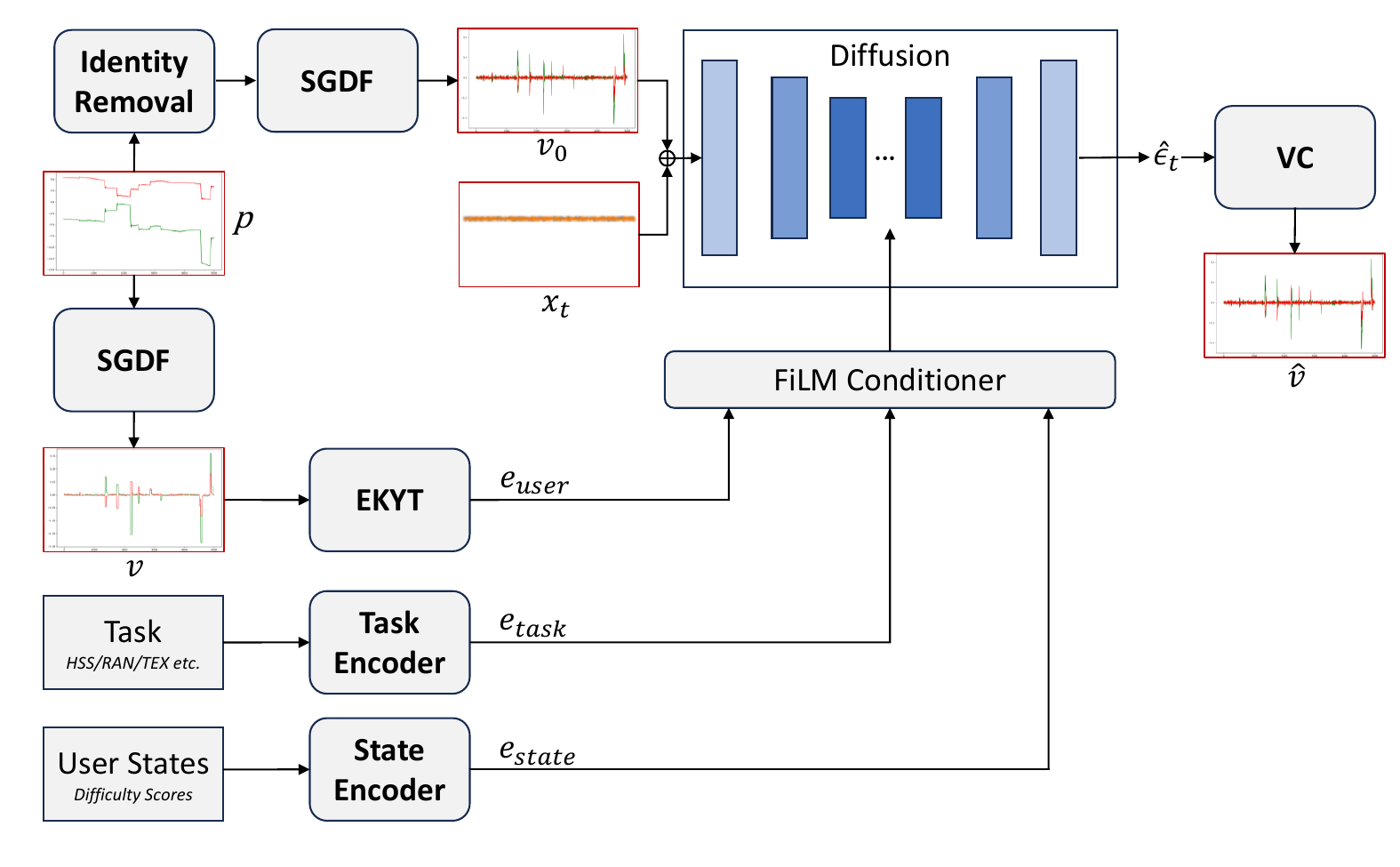}
    \caption{Overview of EyeMakeYou. The real position signal $p$ is processed by identity removal and SGDF to obtain the identity-removed reference velocity $v_0$, while SGDF also produces the real velocity signal $v$ used to extract the user embedding $e_{\mathrm{user}}$ with EKYT. At diffusion step $t$, the noised velocity $x_t$ and $v_0$ are provided to the diffusion model, which is FiLM-conditioned on user, task, and user-state embeddings. The predicted noise $\hat{\epsilon}_t$ is converted by VC into the synthesized velocity signal $\hat{v}$.}
    \label{fig:main_architecture}
\end{figure*}


\section{Proposed Architecture}

\label{sec:EyeMakeYou}
\subsection{Overview}
The proposed multi-conditional diffusion model, EyeMakeYou, is illustrated in Fig.~\ref{fig:main_architecture}. It is a conditional denoising diffusion probabilistic model (DDPM) \cite{ho2020denoising} that synthesizes two-dimensional eye-movement velocity signals conditioned on user identity, task context, and subjective user state. The model operates on 5-s gaze segments sampled at $f_s=1000$~Hz, each segment contains $S=5000$ temporal samples. The DDPM configuration uses 50 diffusion steps, a 30-layer dilated residual denoiser, and three global conditioning sources: an identity representation, a task embedding, and a subjective-state representation.


\subsection{Gaze preprocessing and velocity representation}

Let,
$\mathbf{p} \in \mathbb{R}^{S \times 2}$
denote a two-dimensional gaze-position sequence, where the two channels correspond to horizontal and vertical gaze positions in degrees. We construct a low-frequency identity-removed reference trajectory, denoted by $\mathbf{p}^{\mathrm{co}}$, by downsampling $\mathbf{p}$ to 20 Hz and linearly interpolating it back to 1000 Hz:
\begin{equation}
\mathbf{p}^{\mathrm{co}} = \mathcal{I}_{20}(\mathbf{p}),
\label{eq:low_frequency_reference}
\end{equation}
where $\mathcal{I}_{20}(\cdot)$ denotes the downsampling--interpolation operation. 
This operation produces a low-frequency approximation of the original gaze-position sequence. According to the Percentage of Variance Accounted For (PVAF) analysis \cite{raju2021determining}, the 0–25 Hz band retains nearly all variance in saccade trajectories (i.e., it preserves signal structure) while diminishing higher-frequency components that convey identity features.

The original and reference position signals are converted into velocity signals using a Savitzky--Golay differentiation filter (SGDF) \cite{savitzky1964smoothing} with window length $7$, polynomial order $2$, and first derivative:
\begin{equation}
\begin{split}
\mathbf{v} &= f_s \cdot \mathrm{SGDF}^{(1)}_{7,2}(\mathbf{p}), \\
\mathbf{v}_{\mathrm{o}} &= f_s \cdot \mathrm{SGDF}^{(1)}_{7,2}(\mathbf{p}^{\mathrm{co}}),
\end{split}
\label{eq:velocity_conversion}
\end{equation}
where $\mathbf{v}, \mathbf{v}_{\mathrm{o}} \in \mathbb{R}^{S \times 2}$ are expressed in degrees per second. Invalid values are replaced by zero, and each velocity component is clipped to the interval $[-1000,1000]$ degrees/s.

We then apply a sine-based normalization to both velocity signals:
\begin{equation}
\begin{split}
\mathbf{x}_0 &= \sin\left(\frac{\pi}{2}\frac{\mathbf{v}}{1000}\right), \\
\mathbf{x}_0^{\mathrm{co}} &=
\sin\left(\frac{\pi}{2}\frac{\mathbf{v}_{\mathrm{o}}}{1000}\right),
\end{split}
\label{eq:sine_normalization}
\end{equation}
where $\mathbf{x}_0,\mathbf{x}_0^{\mathrm{co}} \in [-1,1]^{2 \times S}$. Here, $\mathbf{x}_0$ is the clean diffusion target and $\mathbf{x}_0^{\mathrm{co}}$ is the low-frequency identity-removed conditional reference.
The corresponding inverse transformation is as follows:
\begin{equation}
\hat{\mathbf{v}} =
\frac{2 \times 1000}{\pi}
\arcsin\left(
\operatorname{clip}(\hat{\mathbf{x}}_0,-1+\epsilon,1-\epsilon)
\right),
\label{eq:inverse_sine_normalization}
\end{equation}
where $\epsilon=10^{-3}$ is used for numerical stability.

\subsection{Conditional Information}
\label{sec:conditional_information}

EyeMakeYou conditions the denoising process on three complementary global factors: user identity, task, and subjective user state. These conditions have distinct meanings and are encoded separately before being combined in the Feature-wise Linear Modulation (FiLM) layers.

\paragraph{User-identity condition.}
The user-identity feature captures stable, subject-specific characteristics of eye-movement dynamics. We extract this feature from the original velocity signal $\mathbf{v}$ using pre-trained Eye Know You Too (EKYT) encoders \cite{lohr2022ekyt}. Let $\phi_k(\cdot)$ denote the $k$th EKYT encoder. After extracting features with each encoder, the four embeddings are averaged, and then $\ell_2$ normalized as:

\begin{equation}
\bar{\mathbf{e}}_{\mathrm{user}}
=
\frac{1}{4}
\sum_{k=1}^{4}
\phi_k(\mathbf{v}),
\qquad
\mathbf{e}_{\mathrm{user}}
=
\frac{
\bar{\mathbf{e}}_{\mathrm{user}}
}{
\left\|
\bar{\mathbf{e}}_{\mathrm{user}}
\right\|_2
}.
\label{eq:e_user}
\end{equation}
The resulting identity feature is
\begin{equation}
\mathbf{e}_{\mathrm{user}}\in\mathbb{R}^{128}.
\end{equation}
Its role is to guide the generated velocity sequence toward the identity characteristics of the intended user.

\paragraph{Task condition.}
The task feature represents the experimental task associated with the gaze sequence. The model considers all the gaze related tasks:
\begin{equation}
\mathcal{T}=
\{
given\_tasks
\}.
\end{equation}
Each task is mapped to an integer identifier $r\in\{0,\ldots,n\}$ and transformed into a learned task embedding:
\begin{equation}
\mathbf{e}_{\mathrm{task}}
=
\mathrm{Emb}_{\mathrm{task}}(r),
\qquad
\mathbf{e}_{\mathrm{task}}\in\mathbb{R}^{128}.
\label{eq:e_task}
\end{equation}
This condition enables the model to account for task-dependent eye-movement dynamics. In contrast to the identity condition, which specifies \emph{who} the generated signal should resemble, the task condition specifies \emph{which task context} the signal should follow.

\paragraph{User-state condition.}
The user-state feature represents the subjective state reported for a recording. It contains three ratings:
\begin{equation}
\mathbf{q}
=
[
q_{\mathrm{difficulty}},
q_{\mathrm{mental\_tiredness}},
q_{\mathrm{eye\_tiredness}}
]
\in[1,7]^3.
\end{equation}
These ratings are normalized to $[-1,1]$ as follows:
\begin{equation}
\mathbf{s}
=
\frac{\mathbf{q}-4}{3}.
\label{eq:state_normalization}
\end{equation}
Missing, zero, or non-finite ratings are replaced by $1$ before normalization. A recording without subjective ratings is assigned $\mathbf{q}=[1,1,1]$. Then, the normalized state vector is encoded by a two-layer multilayer perceptron as follows:
\begin{equation}
\mathbf{e}_{\mathrm{state}}
=
\rho\,
f_{\mathrm{state}}(\mathbf{s})
=
\rho\left[
\mathbf{W}_{s,2}
\,
\mathrm{SiLU}
\left(
\mathbf{W}_{s,1}\mathbf{s}+\mathbf{b}_{s,1}
\right)
+
\mathbf{b}_{s,2}
\right],
\label{eq:e_state}
\end{equation}
where the hidden and output dimensions are both 32, and $\rho=0.25$ is the user-state conditioning scale. Thus,
\begin{equation}
\mathbf{e}_{\mathrm{state}}\in\mathbb{R}^{32}.
\end{equation}
This condition allows the model to account for recording-specific subjective variation beyond the stable identity feature and the task label.

\paragraph{Combined condition.}
The three features retain their individual meanings until FiLM modulation. They are concatenated into one global condition vector:
\begin{equation}
\mathbf{c}
=
[
\mathbf{e}_{\mathrm{user}};
\mathbf{e}_{\mathrm{task}};
\mathbf{e}_{\mathrm{state}}
]
\in\mathbb{R}^{288}.
\label{eq:combined_condition}
\end{equation}
Therefore, the model receives explicit information about the intended user identity, the performed task, and the user's subjective state.


\subsection{Forward Diffusion Process}

Let $\mathbf{x}_0$ denote the clean normalized velocity signal from Equation~\ref{eq:sine_normalization}. The forward diffusion process gradually adds Gaussian noise as follows:
\begin{equation}
q(\mathbf{x}_t\mid\mathbf{x}_0)
=
\mathcal{N}
\left(
\sqrt{\bar{\alpha}_t}\mathbf{x}_0,
(1-\bar{\alpha}_t)\mathbf{I}
\right),
\end{equation}
Here,
\begin{equation}
\alpha_t=1-\beta_t,
\qquad
\bar{\alpha}_t=\prod_{j=1}^{t}\alpha_j.
\end{equation}
This diffusion model uses $T=50$ diffusion steps and a linear schedule with $\beta_t$ increasing from $10^{-4}$ to $5\times10^{-2}$.

\subsection{Conditional Diffusion Denoiser}

The denoiser predicts the noise added to $\mathbf{x}_0$ as follows:
\begin{equation}
\hat{\boldsymbol{\varepsilon}}_t
=
\varepsilon_{\theta}
\left(
\mathbf{x}_t,
t,
\mathbf{x}_0^{\mathrm{co}},
\mathbf{e}_{\mathrm{user}},
\mathbf{e}_{\mathrm{task}},
\mathbf{e}_{\mathrm{state}}
\right),
\label{eq:conditional_denoiser}
\end{equation}
where $\mathbf{x}_0^{\mathrm{co}}$ is the identity-removed reference velocity signal.

The noisy velocity and reference velocity are concatenated along the channel dimension and projected to a 64-channel temporal representation:
\begin{equation}
\mathbf{h}^{(0)}
=
\mathrm{ReLU}
\left(
\mathrm{Conv}_{1\times1}
\left(
[\mathbf{x}_t;\mathbf{x}_0^{\mathrm{co}}]
\right)
\right)
+
g_{\mathrm{task}}
\,
\mathbf{W}_{\mathrm{task}}
\mathbf{e}_{\mathrm{task}},
\label{eq:denoiser_initial_feature}
\end{equation}
where $g_{\mathrm{task}}$ is a learned scalar initialized to $0.1$. This input-level term provides a direct task-dependent bias before the residual blocks.

The diffusion step $t$ is represented by a 128-dimensional sinusoidal embedding:
\begin{equation}
\mathbf{a}_t =
\left[
\sin\left(t\,10^{4j/63}\right);
\cos\left(t\,10^{4j/63}\right)
\right]_{j=0}^{63}.
\end{equation}
It is transformed through two SiLU-activated linear layers:
\begin{equation}
\mathbf{d}_t
=
\mathrm{SiLU}
\left(
\mathbf{W}_{d,2}
\,
\mathrm{SiLU}
\left(
\mathbf{W}_{d,1}\mathbf{a}_t+\mathbf{b}_{d,1}
\right)
+
\mathbf{b}_{d,2}
\right),
\qquad
\mathbf{d}_t\in\mathbb{R}^{512}.
\label{eq:time_embedding_updated}
\end{equation}

The denoiser contains 30 gated dilated residual blocks with 64 channels. The dilation factors repeat the sequence
\begin{equation}
1,2,4,\ldots,512.
\end{equation}
For residual block $\ell$, the timestep feature is first projected and added to the block input:
\begin{equation}
\mathbf{y}^{(\ell)}
=
\mathrm{DilatedConv}^{(\ell)}
\left(
\mathbf{h}^{(\ell-1)}
+
\mathbf{W}^{(\ell)}_{d}\mathbf{d}_t
\right).
\label{eq:residual_temporal_feature}
\end{equation}

\paragraph{FiLM modulation by identity, task, and user state.}
The three condition embeddings in Equation~\ref{eq:combined_condition} are used jointly to compute the FiLM scale and shift parameters for each residual block:
\begin{equation}
\mathbf{u}^{(\ell)}
=
\mathbf{W}^{(\ell)}_{\mathrm{cond}}\mathbf{c},
\end{equation}
\begin{equation}
[
\boldsymbol{\gamma}^{(\ell)};
\boldsymbol{\beta}^{(\ell)}
]
=
\mathrm{Conv}^{(\ell)}_{\mathrm{affine}}
\left(
\operatorname{repeat}
\left(
\mathbf{u}^{(\ell)},S
\right)
\right).
\label{eq:film_parameters_updated}
\end{equation}
Here, $\boldsymbol{\gamma}^{(\ell)}$ and $\boldsymbol{\beta}^{(\ell)}$ are the learned FiLM scale and shift parameters, respectively. The temporal feature is then modulated as
\begin{equation}
\tilde{\mathbf{y}}^{(\ell)}
=
\mathbf{y}^{(\ell)}
\odot
\left(
1+
\tanh(g^{(\ell)}_{\mathrm{cond}})
\boldsymbol{\gamma}^{(\ell)}
\right)
+
\tanh(g^{(\ell)}_{\mathrm{cond}})
\boldsymbol{\beta}^{(\ell)}.
\label{eq:film_updated}
\end{equation}

Consequently, every residual block receives the same semantically meaningful global condition:
\begin{equation}
[
\underbrace{\mathbf{e}_{\mathrm{user}}}_{\text{identity}};
\underbrace{\mathbf{e}_{\mathrm{task}}}_{\text{task context}};
\underbrace{\mathbf{e}_{\mathrm{state}}}_{\text{subjective state}}
].
\end{equation}
The learned FiLM parameters determine how these three factors scale and shift the temporal features during denoising. Later, the modulated features are processed by a gated activation:
\begin{equation}
[
\mathbf{a}^{(\ell)};
\mathbf{b}^{(\ell)}
]
=
\operatorname{split}
\left(
\tilde{\mathbf{y}}^{(\ell)}
\right),
\qquad
\mathbf{g}^{(\ell)}
=
\sigma
\left(
\mathbf{a}^{(\ell)}
\right)
\odot
\tanh
\left(
\mathbf{b}^{(\ell)}
\right).
\end{equation}
Here, a $1\times1$ convolution produces residual and skip features:
\begin{equation}
[
\mathbf{r}^{(\ell)};
\mathbf{k}^{(\ell)}
]
=
\mathrm{Conv}^{(\ell)}_{\mathrm{out}}
\left(
\mathbf{g}^{(\ell)}
\right),
\end{equation}
and the residual stream is updated by the following formula:
\begin{equation}
\mathbf{h}^{(\ell)}
=
\frac{
\mathbf{h}^{(\ell-1)}
+
\mathbf{r}^{(\ell)}
}{
\sqrt{2}
}.
\end{equation}

Finally, skip features from all residual blocks are combined to predict the diffusion noise:
\begin{equation}
\hat{\boldsymbol{\varepsilon}}_t
=
\mathrm{Conv}_{\mathrm{out}}
\left(
\mathrm{ReLU}
\left(
\mathrm{Conv}_{\mathrm{skip}}
\left(
\frac{1}{\sqrt{30}}
\sum_{\ell=1}^{30}
\mathbf{k}^{(\ell)}
\right)
\right)
\right).
\label{eq:final_noise_prediction}
\end{equation}

\subsection{Reverse Diffusion and Gaze Synthesis}

Given the predicted noise, the clean normalized velocity estimate is achieved by the  velocity-converter (VC) as follows:
\begin{equation}
\hat{\mathbf{x}}_0
=
\frac{
\mathbf{x}_t
-
\sqrt{1-\bar{\alpha}_t}
\hat{\boldsymbol{\varepsilon}}_t
}{
\sqrt{\bar{\alpha}_t}
}.
\label{eq:x0_estimate_updated}
\end{equation}
The reverse-process mean is
\begin{equation}
\boldsymbol{\mu}_{\theta}
=
\frac{1}{\sqrt{\alpha_t}}
\left(
\mathbf{x}_t
-
\frac{1-\alpha_t}
{\sqrt{1-\bar{\alpha}_t}}
\hat{\boldsymbol{\varepsilon}}_t
\right),
\end{equation}
with posterior variance
\begin{equation}
\tilde{\beta}_t
=
\frac{
1-\bar{\alpha}_{t-1}
}{
1-\bar{\alpha}_t
}
\beta_t.
\end{equation}
Starting from Gaussian noise, reverse sampling is performed as
\begin{equation}
\mathbf{x}_{t-1}
=
\boldsymbol{\mu}_{\theta}
+
\sqrt{\tilde{\beta}_t}\boldsymbol{\eta},
\qquad
\boldsymbol{\eta}
\sim
\mathcal{N}(\mathbf{0},\mathbf{I}),
\end{equation}
for $t>1$. The final denoising step is deterministic. The generated normalized velocity is converted to degrees/s using Equation~\ref{eq:inverse_sine_normalization} and integrated from the initial seed position to obtain the synthetic gaze-position trajectory.

\subsection{Loss Functions}
\label{sec:loss_functions}

\subsubsection{Noise-Prediction Loss}

The primary diffusion objective is mean-squared error between the true and predicted Gaussian noise:
\begin{equation}
\mathcal{L}_{\mathrm{noise}}
=
\frac{1}{2BS}
\sum_{b=1}^{B}
\sum_{c=1}^{2}
\sum_{s=1}^{S}
\left(
\hat{\varepsilon}_{b,c,s}
-
\varepsilon_{b,c,s}
\right)^2.
\end{equation}

\subsubsection{Identity-Preservation Loss}

The identity loss aligns the EKYT representation of generated velocity with that of the real velocity:
\begin{equation}
\mathcal{L}_{\mathrm{id}}
=
\frac{1}{B}
\sum_{b=1}^{B}
\left[
1-
\frac{
\phi(\mathbf{v}_b)^{\top}
\phi(\hat{\mathbf{v}}_b)
}{
\left\|
\phi(\mathbf{v}_b)
\right\|_2
\left\|
\phi(\hat{\mathbf{v}}_b)
\right\|_2
}
\right].
\label{eq:identity_loss_updated}
\end{equation}

\subsubsection{Proposed Signal-Fidelity Losses}
\label{sec:proposed_losses}

In addition to diffusion noise prediction and identity preservation, we introduce three complementary signal-fidelity objectives: multi-resolution spectral loss, drift-consistency loss, and event-weighted local smoothness loss. These losses act on different properties of synthesized eye-movement velocity signals. Specifically, the spectral loss constrains frequency-domain structure, the drift loss constrains cumulative displacement after integration, and the local smoothness loss constrains short-term temporal dynamics in event-relevant regions.

\subsubsection{Multi-Resolution Spectral Loss}

The diffusion noise-prediction objective reconstructs the clean velocity signal indirectly. However, minimizing pointwise reconstruction error alone does not explicitly ensure that the generated signal reproduces the frequency characteristics of real eye movements. We therefore compare the short-time Fourier transform (STFT) magnitudes of the generated and real normalized velocity signals at multiple temporal resolutions.

Let $\mathbf{x}_0$ and $\hat{\mathbf{x}}_0$ denote the real and generated normalized velocity signals, respectively. For each FFT size
\begin{equation}
n \in \mathcal{N}_{\mathrm{FFT}} =
\{64,128,256,512\},
\end{equation}
we compute a Hann-windowed STFT with hop size
\begin{equation}
h_n = \operatorname{round}(0.25n).
\end{equation}
The magnitude spectrum is defined as
\begin{equation}
\mathcal{M}_n(\mathbf{x})
=
\sqrt{
\Re\left(\mathrm{STFT}_n(\mathbf{x})\right)^2
+
\Im\left(\mathrm{STFT}_n(\mathbf{x})\right)^2
+
\epsilon
},
\end{equation}
where $\epsilon=10^{-7}$ ensures numerical stability. The multi-resolution spectral loss is
\begin{equation}
\mathcal{L}_{\mathrm{spec}}
=
\frac{1}{|\mathcal{N}_{\mathrm{FFT}}|}
\sum_{n\in\mathcal{N}_{\mathrm{FFT}}}
\left\|
\mathcal{M}_n(\hat{\mathbf{x}}_0)
-
\mathcal{M}_n(\mathbf{x}_0)
\right\|_1.
\label{eq:multiresolution_spectral_loss}
\end{equation}

Using several FFT sizes constrains spectral structure at different temporal scales. The shorter windows emphasize rapid local changes, whereas longer windows constrain broader temporal frequency patterns. Consequently, this loss encourages the generated velocity signal to match the multi-scale frequency characteristics of the real signal rather than only its sample-wise values.

\subsubsection{Drift-Consistency Loss}

Gaze positions are obtained by integrating velocity over time. Therefore, even a small persistent velocity bias can accumulate into a substantial spatial displacement error. To reduce this accumulated error, we introduce a drift-consistency loss that compares the final integrated displacement of the generated and real velocity signals.

Let $\mathbf{v}_{b,s}\in\mathbb{R}^{2}$ and $\hat{\mathbf{v}}_{b,s}\in\mathbb{R}^{2}$ denote the real and generated physical velocities in degrees/s for sample $s$ of batch element $b$, respectively. We define a validity mask $m_{b,s}$ that excludes non-finite samples and samples labeled as Basic Noise or Event Detection Noise:
\begin{equation}
m_{b,s}
=
\mathbb{I}
\left[
\mathbf{v}_{b,s}
\text{ and }
\hat{\mathbf{v}}_{b,s}
\text{ are finite}
\right]
\mathbb{I}[a_{b,s}\neq4]
\mathbb{I}[a_{b,s}\neq5],
\end{equation}
where $a_{b,s}$ denotes the available event annotation.

The real and generated displacement vectors are
\begin{equation}
\mathbf{d}_b
=
\frac{1}{f_s}
\sum_{s=1}^{S}
m_{b,s}\mathbf{v}_{b,s},
\qquad
\hat{\mathbf{d}}_b
=
\frac{1}{f_s}
\sum_{s=1}^{S}
m_{b,s}\hat{\mathbf{v}}_{b,s}.
\end{equation}
The drift-consistency loss is
\begin{equation}
\mathcal{L}_{\mathrm{drift}}
=
\frac{1}{2B}
\sum_{b=1}^{B}
\left\|
\hat{\mathbf{d}}_b-\mathbf{d}_b
\right\|_1.
\label{eq:drift_consistency_loss}
\end{equation}

This objective directly constrains cumulative velocity error. As a result, it reduces the tendency of small systematic velocity deviations to produce large position drift after temporal integration. The noise-related annotations are used only to exclude unreliable samples from this displacement comparison; they are not provided to the denoiser as conditioning inputs.

\subsubsection{Event-Weighted Local Smoothness Loss}

Eye-movement velocity has different local temporal characteristics across movement events. In particular, fixation regions should have stable local dynamics, whereas saccades should not be excessively smoothed. We therefore introduce an event-weighted local smoothness loss that compares the second-order temporal differences of generated and real normalized velocities.

For a normalized velocity signal $\mathbf{x}$, the local second-order temporal difference is
\begin{equation}
\Delta^2\mathbf{x}_{b,:,s}
=
\mathbf{x}_{b,:,s+1}
-
2\mathbf{x}_{b,:,s}
+
\mathbf{x}_{b,:,s-1}.
\label{eq:second_order_difference}
\end{equation}
We assign event-dependent weights using the annotation at the central sample:
\begin{equation}
w_{b,s}
=
\begin{cases}
1, & a_{b,s}=\mathrm{Fixation},\\
0.25, & a_{b,s}=\mathrm{PSO},\\
0, & \text{otherwise}.
\end{cases}
\label{eq:event_weights}
\end{equation}
The event-weighted local smoothness loss is
\begin{equation}
\mathcal{L}_{\mathrm{smooth}}
=
\frac{
\displaystyle
\sum_{b=1}^{B}
\sum_{c=1}^{2}
\sum_{s=2}^{S-1}
w_{b,s}
\left|
\Delta^2\hat{x}_{0,b,c,s}
-
\Delta^2x_{0,b,c,s}
\right|
}{
\displaystyle
2
\sum_{b=1}^{B}
\sum_{s=2}^{S-1}
w_{b,s}
}.
\label{eq:event_weighted_smoothness_loss}
\end{equation}
If no fixation or PSO samples occur in a batch, $\mathcal{L}_{\mathrm{smooth}}$ is set to zero.

This loss does not simply minimize the roughness of the generated signal. Instead, it matches generated local curvature to the real velocity signal, primarily within fixation intervals and, to a lesser extent, within post-saccadic oscillation intervals. By assigning zero weight to saccades and noise samples, the loss avoids imposing fixation-like smoothness on rapid eye movements while suppressing inappropriate local fluctuations in regions expected to be relatively stable.

\subsubsection{Combined Objective}

The three proposed losses are combined with the diffusion noise-prediction and identity-preservation losses:
\begin{equation}
\mathcal{L}_{\mathrm{total}}
=
\mathcal{L}_{\mathrm{noise}}
+
0.5\mathcal{L}_{\mathrm{id}}
+
0.1\mathcal{L}_{\mathrm{spec}}
+
0.003\mathcal{L}_{\mathrm{drift}}
+
0.001\mathcal{L}_{\mathrm{smooth}}.
\label{eq:combined_proposed_loss}
\end{equation}

The three proposed terms are complementary: $\mathcal{L}_{\mathrm{spec}}$ preserves multi-scale frequency structure, $\mathcal{L}_{\mathrm{drift}}$ preserves integrated displacement, and $\mathcal{L}_{\mathrm{smooth}}$ preserves event-relevant local temporal dynamics.


\section{Experiments}
\subsection{Dataset}
For our experiment, we used the publicly available GazeBase \cite{griffith2021gazebase} dataset to train and evaluate performance. This dataset was collected using the Eyelink 1000 eye tracker and captured monocular (i.e., left-eye) movement data at 1,000 Hz. Nine rounds were conducted over 37 months, involving 322 volunteers and yielding 12,334 eye-tracking records. 
Specifically, it contains seven tasks: fixation (FXS), horizontal saccade (HSS), random oblique saccade (RAN), reading (TEX), free viewing of cinematic video (VD1 and VD2), and gaze-driven gaming (BLG). Furthermore, among these seven tasks, only RAN and HSS have corresponding stimulus signals that enable the calculation of spatial accuracy and spatial precision metrics. 
Refer to the GazeBase \cite{griffith2021gazebase} paper for further information. In particular, the training set comprises recordings from the 263 participants. The test set includes recordings from 59 subjects who were present in round 6, and synthetic 5-second sequences are concatenated in temporal order to match the corresponding real recording length, enabling direct synthetic-real comparison. There was no overlap between users in the training and test sets, ensuring that evaluations were performed on completely unseen users.

\subsection{Implementation Details}
\label{sec:implementation_details}

EyeMakeYou was implemented in PyTorch 2.1.0. The model receives a four-channel temporal input formed by concatenating the noisy normalized velocity signal and the low-frequency reference velocity signal. The denoiser contains 30 gated dilated residual blocks with 64 residual channels and a dilation cycle length of 10. User identity was represented using a 128-dimensional embedding obtained by averaging the outputs of four frozen EKYT models. A learned 128-dimensional task embedding represented each task. A two-layer multilayer perceptron encoded the three-dimensional subjective user-state vector into a 32-dimensional state embedding. The identity, task, and user-state embeddings were concatenated and injected into every residual block through FiLM modulation. In addition, the task embedding was projected to the residual channel dimension and added as a global, task-dependent bias at the denoiser input—the reported configuration used a user-state conditioning scale of $0.25$. Training was performed on an NVIDIA RTX A6000 GPU. On CUDA-enabled hardware, automatic mixed precision with \texttt{bfloat16} autocasting was used. TensorFloat-32 computation was enabled for CUDA matrix multiplication and cuDNN operations. To ensure the reproducibility of our work, we will release the complete source code and trained models.

\subsection{Training Details}
\label{sec:training_details}

Training samples consisted of non-overlapping 5-s gaze windows sampled at 1000 Hz, corresponding to 5000 samples per window. For each window, raw gaze positions and the 20-Hz low-frequency reference positions were converted to velocity using a Savitzky--Golay differentiation filter with window length $7$, polynomial order $2$, and first derivative. Velocities were clipped to $[-1000,1000]$ degrees/s and normalized to $[-1,1]$ using the sine-based transformation described in Section~\ref{sec:EyeMakeYou}.

We trained the model using a DDPM process with $T=50$ diffusion steps. The forward process used a linear noise schedule with $\beta_t$ increasing from $10^{-4}$ to $5\times10^{-2}$. At each training iteration, a diffusion step was sampled uniformly, Gaussian noise was added to the clean normalized velocity, and the model was optimized to predict that noise.

The model was optimized using AdamW \cite{loshchilov2017decoupled} with a learning rate of $2\times10^{-4}$ and a batch size of 32. Gradient norms were clipped to a maximum value of $1.0$. Training was performed for 300 epochs with a fixed random seed of 1337. Model checkpoints were saved every 50 epochs.

The training objective combined diffusion noise prediction, EKYT-based identity preservation, multi-resolution spectral similarity, drift consistency, and event-weighted local smoothness:
\begin{equation}
\mathcal{L}_{\mathrm{total}}
=
\mathcal{L}_{\mathrm{noise}}
+
0.5\mathcal{L}_{\mathrm{id}}
+
0.1\mathcal{L}_{\mathrm{spec}}
+
0.003\mathcal{L}_{\mathrm{drift}}
+
0.001\mathcal{L}_{\mathrm{smooth}}.
\end{equation}


\subsection{Evaluation Metrics}
\label{sec:evaluation-metrics}

We evaluate the proposed synthetic gaze generator at four complementary levels: spatial signal quality, subject-specific embedding similarity, qualitative comparisons, and preservation of physiologically meaningful oculomotor characteristics.


\paragraph{Spatial accuracy and precision.}
We evaluate spatial accuracy and spatial precision during stable fixation periods. Specifically, we identify 80-ms fixation bins prior to measurement \cite{lohr2019evaluating}.

For a synthetic gaze vector $\hat{\mathbf{g}}$ and the corresponding eye-to-target vector $\mathbf{t}$, spatial accuracy is the angular error, in degrees of visual angle (dva):
\begin{equation}
\theta =
\frac{180}{\pi}
\cos^{-1}
\left(
\frac{\hat{\mathbf{g}} \cdot \mathbf{t}}
{\left\| \hat{\mathbf{g}} \right\|
 \left\| \mathbf{t} \right\|}
\right).
\label{eq:spatial-accuracy}
\end{equation}

Spatial precision measures sample-to-sample instability within a stable fixation bin. For user $u$, bin $b$, and $n_b$ samples, we compute root-mean-square (RMS) precision as
\begin{equation}
\rho_{u,b} =
\sqrt{
\frac{1}{n_b-1}
\sum_{k=2}^{n_b}
\left\|
\hat{\mathbf{g}}_{u,b,k}
-
\hat{\mathbf{g}}_{u,b,k-1}
\right\|_2^2
}.
\label{eq:spatial-precision}
\end{equation}
where lower values indicate better spatial accuracy and precision.

Following the user-centric evaluation framework of \cite{aziz2024evaluation}, we report both error percentiles ($E$) and user percentiles ($U$). For each user, $E50$ and $E95$ denote the median and 95th-percentile error, respectively, across that user's stable fixation bins. The $U50$ and $U95$ users represent the median and 95th percentiles of average signal quality across the user population. Thus, $U50|E50$ represents an average user under average-case samples, whereas $U95|E95$ represents the error level required to support 95\% of users for 95\% of their samples.

\paragraph{Synthetic--real embedding similarity.}
To evaluate preservation of subject-specific gaze dynamics, we extract embeddings from real and synthetic signals using the same pretrained eye-movement embedding model. For $N$ matched real--synthetic sequence pairs with embeddings $\mathbf{z}_i^R$ and $\mathbf{z}_i^S$, respectively, we report mean cosine similarity:
\begin{equation}
\mathrm{CosSim}
=
\frac{1}{N}
\sum_{i=1}^{N}
\frac{
\mathbf{z}_i^R \cdot \mathbf{z}_i^S
}{
\left\|\mathbf{z}_i^R\right\|_2
\left\|\mathbf{z}_i^S\right\|_2
}.
\label{eq:embedding-similarity}
\end{equation}
where higher cosine similarity indicates that synthetic signals preserve the identity-relevant dynamics of their corresponding real signals.



\section{Results}
\subsection{Spatial Accuracy}

Tab.~\ref{tab:spatial_accuracy} reports the U$\mid$E spatial-accuracy results for the HSS and RAN tasks, where lower gaze-point error in dva indicates better agreement with the target stimulus position. At the median user and error level (U50$\mid$E50), the proposed EyeMakeYou achieves the lowest errors for both tasks, with 3.60~dva for HSS and 2.81~dva for RAN. These values substantially improve upon SP-EyeGAN (15.45 and 13.63~dva), the VAE baseline (18.65 and 13.50~dva), and DiffEyeSyn (3.73 and 4.06~dva), respectively.

Under the more demanding U95$\mid$E95 condition, EyeMakeYou achieves the lowest HSS error (31.15~dva), closely approaching the ground-truth value of 30.85~dva and improving upon DiffEyeSyn (38.61~dva), SP-EyeGAN (52.73~dva), and the VAE (50.81~dva). For RAN, EyeMakeYou achieves 28.77~dva, substantially improving upon SP-EyeGAN (47.07~dva) and the VAE (34.75~dva), although DiffEyeSyn obtains a slightly lower error of 26.77~dva. Overall, EyeMakeYou provides the strongest median spatial accuracy across both tasks and robust performance under difficult HSS conditions. Although it does not fully match ground-truth accuracy, particularly at higher percentiles, the results demonstrate that the proposed method generates gaze signals that remain substantially closer to the target stimulus than the GAN- and VAE-based baselines across both typical and challenging user conditions.

From an application perspective, spatial accuracy is particularly important for gaze-based pointing, dwell-based target selection, and foveated-rendering systems, where positional deviations directly affect usability. The low U50$\mid$E50 errors of EyeMakeYou suggest that synthetic data generated by the proposed method can better support the development and evaluation of interfaces designed for typical users, potentially enabling smaller target regions or tighter interaction margins than data generated by SP-EyeGAN or the VAE. Nevertheless, the U95$\mid$E95 results show that substantial errors remain for challenging users and sequences, especially for RAN. Therefore, applications trained or evaluated with synthetic gaze data should retain conservative target-sizing policies, adaptive calibration procedures, or uncertainty-aware interaction mechanisms to maintain reliable performance under worst-case conditions.

\begin{table}[h]
\centering
\caption{U\textbar{}E spatial accuracy for HSS and RAN. Ground truth denotes the real positional signal; $\downarrow$ indicates lower is better.}
\label{tab:spatial_accuracy}
\resizebox{0.4\columnwidth}{!}{%
\begin{tabular}{lcccc}
\toprule
\multirow{2}{*}{Model} & \multicolumn{2}{c}{U50\textbar{}E50 $\downarrow$} & \multicolumn{2}{c}{U95\textbar{}E95 $\downarrow$} \\
\cmidrule(lr){2-3} \cmidrule(lr){4-5}
 & HSS & RAN & HSS & RAN \\
\midrule
Ground truth & 0.89  & 0.92  & 30.85 & 21.80 \\
SP-EyeGAN    & 15.45 & 13.63 & 52.73 & 47.07 \\
VAE          & 18.65  & 13.50  & 50.81  & 34.75 \\
DiffEyeSyn   & 3.73  & 4.06  & 38.61 & 26.77 \\
Ours   & 3.60  & 2.81  & 31.15  & 28.77 \\
\bottomrule
\end{tabular}}
\end{table}

\subsection{Spatial Precision}

Tab.~\ref{tab:spatial_precision} presents the U$\mid$E spatial-precision results, measured as RMS gaze dispersion during steady gaze behavior; lower values indicate less jitter. At the U50$\mid$E50 level, all synthetic approaches obtain values of 0.01~degrees RMS for both HSS and RAN, indicating similar median fixation stability and providing little discrimination among the models under typical conditions. Differences become more apparent at U95$\mid$E95. The proposed EyeMakeYou achieves the lowest worst-case precision error for HSS (0.21~degrees RMS), improving upon DiffEyeSyn (0.35~degrees RMS), SP-EyeGAN (0.37~degrees RMS), and the VAE baseline (0.23~degrees RMS). For RAN, EyeMakeYou obtains 0.29~degrees RMS, matching DiffEyeSyn and improving upon SP-EyeGAN (0.41~degrees RMS), although the VAE achieves the lowest value of 0.17~degrees RMS. Importantly, all synthetic methods exhibit considerably lower U95$\mid$E95 dispersion than the ground-truth signals. Although lower RMS dispersion is desirable from a stability perspective, this pattern may also indicate that generative models smooth some high-variability behavior observed in real gaze data. Thus, precision should be interpreted jointly with spatial accuracy and biometric-preservation results.

Spatial precision is critical for applications involving fixation analysis, gaze-contingent rendering, dwell-time interaction, reading assessment, and selection of small visual targets. The reduced worst-case jitter of EyeMakeYou, particularly for HSS, is beneficial for such applications because it decreases the likelihood that synthetic trajectories introduce unstable gaze points or abrupt fixation noise. In practice, this can support more reliable simulation and augmentation of gaze-driven interfaces, particularly for systems that depend on stable fixation estimates. However, because the synthetic signals may suppress part of the real high-percentile variability, EyeMakeYou-generated data can be used primarily to augment training and evaluation datasets rather than as a complete replacement for real gaze recordings in applications requiring precise calibration or modeling of extreme user behavior.

\begin{table}[h]
\centering
\caption{U\textbar{}E spatial precision for HSS and RAN. Ground truth denotes the real positional signal; $\downarrow$ indicates lower is better.}
\label{tab:spatial_precision}
\resizebox{0.4\columnwidth}{!}{%
\begin{tabular}{lcccc}
\toprule
\multirow{2}{*}{Model} & \multicolumn{2}{c}{U50\textbar{}E50 $\downarrow$} & \multicolumn{2}{c}{U95\textbar{}E95 $\downarrow$} \\
\cmidrule(lr){2-3} \cmidrule(lr){4-5}
 & HSS & RAN & HSS & RAN \\
\midrule
Ground truth & 0.01 & 0.01 & 0.86 & 1.14 \\
SP-EyeGAN    & 0.01 & 0.01 & 0.37 & 0.41 \\
VAE          & 0.01 & 0.01 & 0.23 & 0.17 \\
DiffEyeSyn   & 0.01 & 0.01 & 0.35 & 0.29 \\
Ours   & 0.01 & 0.01 & 0.21 & 0.29 \\
\bottomrule
\end{tabular}}
\end{table}

\subsection{Synthetic Data Similarity}

To assess whether the generated sequences preserve subject-specific biometric characteristics, we measured cosine similarity between embeddings extracted from real and synthetic gaze signals using the pre-trained EKYT model. Higher cosine similarity indicates that the synthetic and real signals occupy more similar locations in the learned eye-movement biometric feature space. As shown in Tab.~\ref{tab:cosine_similarity}, the proposed EyeMakeYou achieves the highest similarity across all seven tasks, with mean values ranging from 0.91 for FXS to 0.95 for BLG and HSS. In contrast, SP-EyeGAN produces substantially lower similarity scores across all tasks, ranging from 0.11 to 0.16, indicating limited preservation of the biometric features captured by EKYT.

EyeMakeYou also outperforms both DiffEyeSyn and the VAE baseline for every task. For example, EyeMakeYou achieves similarities of $0.95 \pm 0.03$, $0.95 \pm 0.01$, and $0.94 \pm 0.01$ for BLG, HSS, and RAN, respectively, compared with $0.92 \pm 0.04$, $0.91 \pm 0.03$, and $0.89 \pm 0.04$ for DiffEyeSyn, and $0.78 \pm 0.04$, $0.73 \pm 0.02$, and $0.71 \pm 0.03$ for the VAE. Overall, these findings indicate that EyeMakeYou most effectively preserves the identity-related eye-movement characteristics of real signals across task types involving fixation, reading, saccadic behavior, video viewing, and gaze-driven interaction.

From an application perspective, high synthetic--real embedding similarity is particularly relevant to gaze-based authentication, where preserving subject-specific behavioral characteristics is more important than generating visually plausible trajectories alone. The consistently high EKYT similarity of EyeMakeYou suggests that its generated signals may be useful for augmenting limited biometric datasets, evaluating authentication pipelines, and simulating user-specific gaze behavior in AR/VR environments. In particular, the strong performance across all tasks indicates that the model is not restricted to a single viewing behavior, which is important for practical systems in which gaze patterns vary with user activity and visual context. Nevertheless, this metric should be interpreted as biometric feature-space consistency according to EKYT; it should be considered together with spatial accuracy, precision, and other signal-level evaluations when assessing overall synthetic-gaze realism.

\begin{table*}[h]
\centering
\caption{Cosine similarity between pre-trained EKYT embeddings of real and synthetic eye movement signals for the evaluated generative models.}
\label{tab:cosine_similarity}
\resizebox{1.0\textwidth}{!}{
\begin{tabular}{lccccccc}
\toprule
\multirow{2}{*}{Model} & \multicolumn{7}{c}{Task} \\
\cmidrule(lr){2-8}
 & BLG & FXS & HSS & RAN & TEX & VD1 & VD2 \\
\midrule
SP-EyeGAN  & $0.14 \pm 0.13$ & $0.11 \pm 0.15$ & $0.11 \pm 0.14$ & $0.13 \pm 0.14$ & $0.16 \pm 0.14$ & $0.13 \pm 0.15$ & $0.12 \pm 0.14$ \\
VAE         & $0.78 \pm 0.04$ & $0.64 \pm 0.06$ & $0.73 \pm 0.02$ & $0.71 \pm 0.03$ & $0.73 \pm 0.03$ & $0.73 \pm 0.04$ & $0.73 \pm 0.04$ \\
DiffEyeSyn & $0.92 \pm 0.04$ & $0.86 \pm 0.05$ & $0.91 \pm 0.03$ & $0.89 \pm 0.04$ & $0.89 \pm 0.03$ & $0.90 \pm 0.04$ & $0.90 \pm 0.03$ \\
Ours  & $0.95 \pm 0.03$ & $0.91 \pm 0.04$ & $0.95 \pm 0.01$ & $0.94 \pm 0.01$ & $0.94 \pm 0.02$ & $0.94 \pm 0.02$ & $0.94 \pm 0.02$ \\
\bottomrule
\end{tabular}
}
\end{table*}

\begin{figure*}[h]
    \centering
    \includegraphics[width=\textwidth]{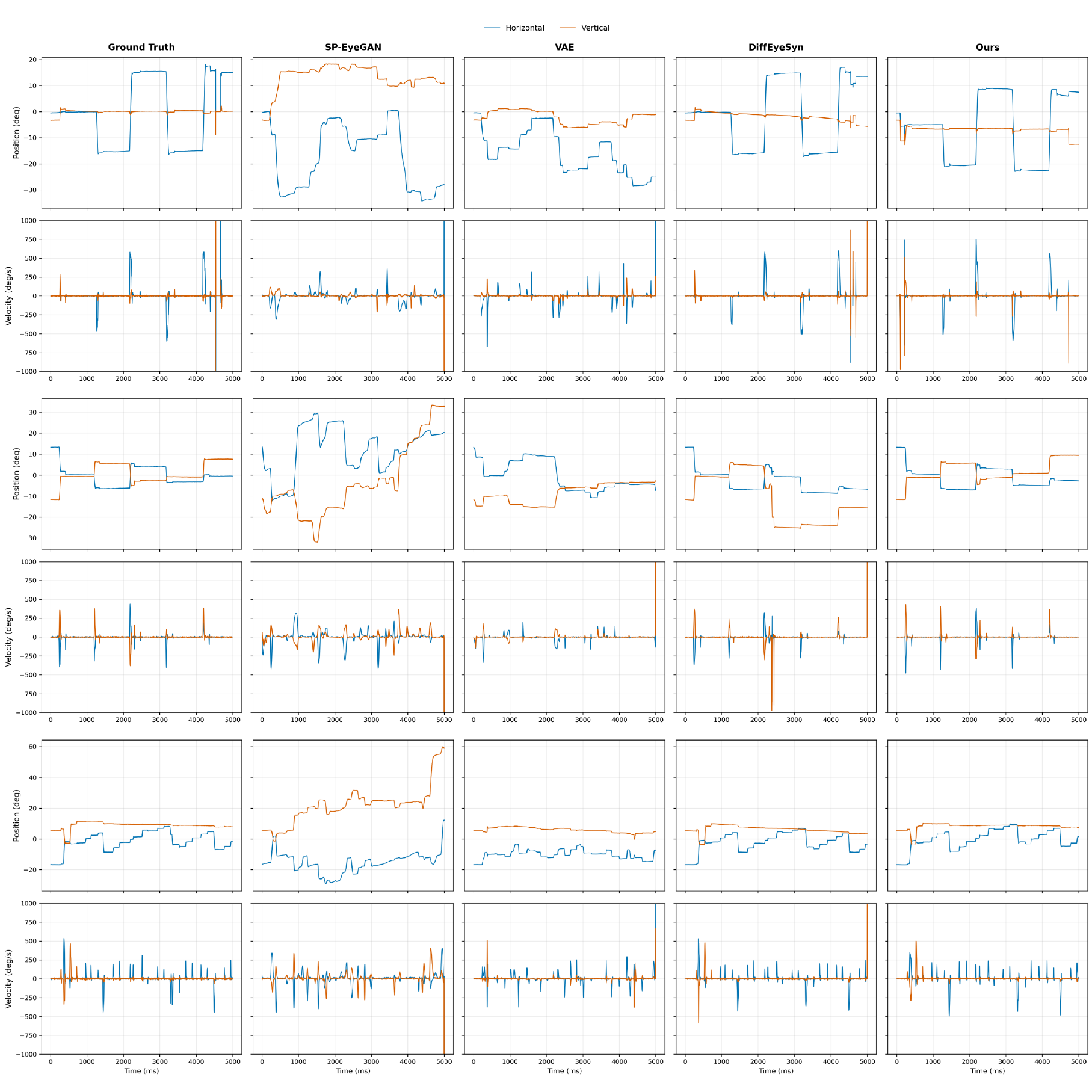}
    \caption{\textbf{Qualitative comparison of real and synthetic eye-movement signals across three tasks.} Columns show Ground Truth, SP-EyeGAN, VAE, DiffEyeSyn, and EyeMakeYou, respectively. The first two rows correspond to the HSS task, the middle two rows correspond to RAN, and the final two rows correspond to TEX. Within each task pair, the upper row shows gaze position and the lower row shows gaze velocity. Each subplot contains horizontal (blue) and vertical (orange) components over a 5-s segment. Velocity is displayed within the range of $[-1000, 1000]$ deg/s for visual comparability.}
    \label{fig:qualitative_examples}
\end{figure*}

\begin{figure*}[!htbp]
    \centering
    \includegraphics[width=\textwidth]{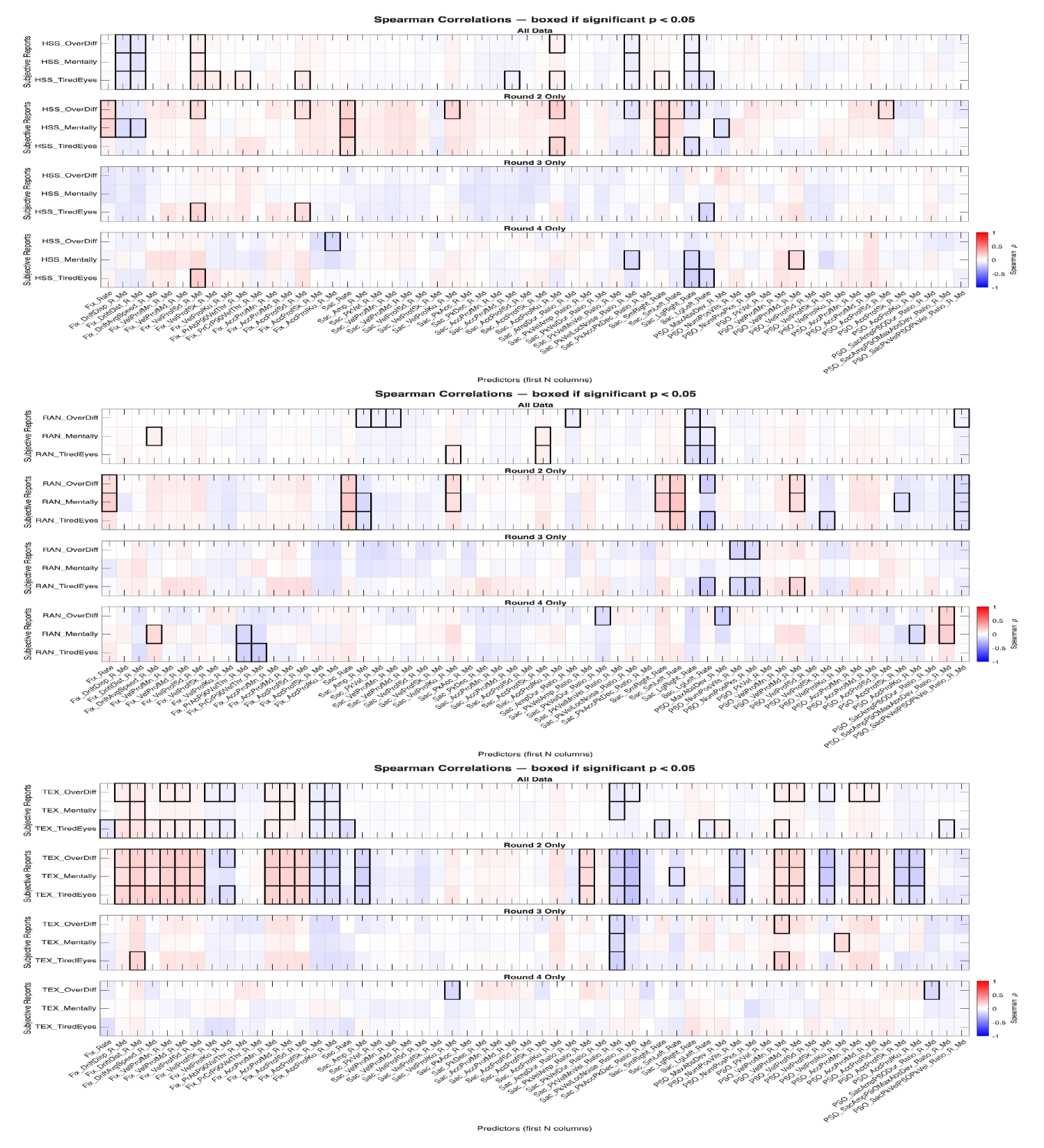}
    \caption{Spearman correlations between subjective reports and eye-movement features for real and synthetic gaze signals. The three task blocks correspond to HSS (top), RAN (middle), and TEX (bottom). Within each task, correlations are shown for pooled data (\emph{All Data}) and separately for Rounds~2--4. Columns correspond to the 58 extracted eye-movement features, and rows correspond to subjective reports. Red and blue indicate positive and negative Spearman correlations, respectively; black boxes denote statistically significant correlations ($p < 0.05$).}
    \label{fig:user_state_correlations}
\end{figure*}

\subsection{Qualitative Evaluations}

Fig. \ref{fig:qualitative_examples} qualitatively compares the real and synthetic bivariate gaze signals produced by SP-EyeGAN, VAE, DiffEyeSyn, and our proposed EyeMakeYou model for the HSS, RAN, and TEX tasks. For each task, the first row shows horizontal (left) and vertical (right) gaze positions, whereas the second row shows the corresponding velocities. Although all models generate visually plausible gaze sequences, clear differences are evident in their preservation of fixation plateaus, saccadic transitions, and velocity dynamics. SP-EyeGAN often produces excessive or misplaced position excursions, unstable fixation segments, and high-frequency velocity jitter. The VAE outputs are generally smoother but tend to attenuate the magnitude and sharpness of gaze transitions, thereby underrepresenting the dynamic structure of saccades. DiffEyeSyn better captures several major position changes; however, it still exhibits occasional misaligned transitions and isolated, saturated velocity spikes that are not consistently supported by the corresponding position trajectories.

Across all three tasks, EyeMakeYou more closely reproduces the qualitative structure of the real signals. In HSS and RAN, it preserves clearer step-like fixation plateaus and more realistic changes in both horizontal and vertical gaze position. In TEX, EyeMakeYou maintains the smaller-amplitude, denser movement patterns without the excessive drift or oscillation observed in some baseline outputs. Its velocity traces also show sparse, localized peaks that correspond more naturally to position transitions, with reduced jitter relative to SP-EyeGAN and fewer abrupt outliers than DiffEyeSyn. Overall, these examples indicate that EyeMakeYou better preserves the task-dependent spatial and temporal characteristics of natural eye movements, producing synthetic signals that are visually more consistent with real oculomotor behavior.

\subsection{User States}

To investigate whether the synthetic gaze signals retain relationships between eye movements and subjective user states, we followed the correlation-based evaluation framework in \cite{qian2025we}. We applied the event-segmentation algorithm in \cite{rigas2018study} and the feature-extraction pipeline in \cite{friedman_novel_2018,friedman_eye_2022} to the synthetic gaze sequences. This procedure yielded 58 interpretable oculomotor features describing fixation, saccade, and drift behavior, including fixation rate, saccade frequency and amplitude, fixation stability, and drift-related measures. Each feature was represented by its event rate or, for multi-valued measures, its median value. For each task, we computed Spearman rank correlations \cite{zar2005spearman} between each subjective-state report and each oculomotor feature across participants. Spearman correlation was used because the subjective ratings were non-normally distributed and exhibited floor effects. Fig. \ref{fig:user_state_correlations} presents the resulting correlations for pooled data and separately for Rounds~2--4; boxed cells indicate statistically significant associations ($p < 0.05$).

The synthetic signals exhibit task- and round-dependent associations between subjective states and oculomotor behavior. For HSS, several significant associations are observed in the pooled analysis and in Round~2, including relationships between subjective reports and fixation-related features. These results indicate that variation in the generated fixation behavior is associated with participants' reported difficulty, mental state, and tired-eye symptoms. However, the HSS associations are relatively sparse and show limited consistency across rounds.

For RAN, significant associations are also observed between subjective reports and a subset of fixation- and saccade-related features. These associations are most apparent in the pooled analysis and Round~2, whereas the patterns in Rounds~3 and~4 are more limited and differ in both direction and feature composition. Thus, the synthetic RAN signals retain some state-related variation, but the relationships are not uniformly stable across experimental rounds.

TEX shows the most numerous and distributed significant associations among the three tasks, particularly in the pooled analysis and Round~2. Subjective reports in this task are associated with multiple fixation-, saccade-, and post-saccadic-oscillation-related features, including fixation-rate measures. The presence of both positive and negative correlations indicates that different aspects of the synthetic eye-movement behavior vary differently with subjective state. In contrast, fewer significant associations are observed in Rounds~3 and~4.

Overall, these results show that the synthetic gaze signals contain detectable relationships between oculomotor features and subjective user states. The associations are strongest and most broadly distributed for TEX, while HSS and RAN show sparser and more round-dependent patterns. Because the analysis evaluates many feature--report pairs, these findings should be interpreted as exploratory correlational evidence; they do not establish causal relationships between subjective state and eye-movement behavior.



\section{Discussion}

This study investigates whether high-frequency synthetic gaze can retain the spatial, temporal, and identity-related characteristics required for eye-movement biometric applications when generation is conditioned on user identity, task, and subjective state. Across the reported evaluations, EyeMakeYou provides a stronger overall trade-off than the GAN and VAE baselines and improves upon DiffEyeSyn in several key settings. Its advantage is most apparent for typical users and samples, where it achieves the lowest median spatial errors for both HSS and RAN and the highest real--synthetic EKYT embedding similarity across all seven tasks. The results nevertheless show that synthetic-gaze quality is multidimensional: low positional error, low fixation jitter, identity-feature preservation, and physiological feature fidelity should not be interpreted as interchangeable measures. A model can improve one aspect of the signal while smoothing, attenuating, or otherwise altering another.

The spatial-accuracy analysis illustrates this distinction. EyeMakeYou achieves the best U50$\mid$E50 accuracy for HSS and RAN, with errors of 3.60 and 2.81~dva, respectively, indicating that it more closely reproduces target-relevant gaze behavior for an average user under typical samples. Its HSS U95$\mid$E95 error of 31.15~dva is also close to the real-signal value of 30.85~dva and lower than the corresponding values for SP-EyeGAN, the VAE, and DiffEyeSyn. This result is encouraging because HSS is dominated by discrete, task-defined horizontal shifts, for which preserving fixation locations and transition timing is particularly important. In RAN, EyeMakeYou again achieves the lowest median error, but its U95$\mid$E95 error remains higher than that of DiffEyeSyn (28.77 versus 26.77~dva). Thus, the proposed conditioning improves typical RAN behavior but does not uniformly improve the most difficult user--sample combinations. This task-dependent pattern supports reporting user-centric operating points rather than relying only on mean performance: a model suitable for average-case data augmentation may still require caution when used to model the tail of the user population.

The spatial-precision results further indicate that low synthetic jitter should not automatically be regarded as complete realism. All synthetic models achieve 0.01~degrees RMS at U50$\mid$E50, and EyeMakeYou attains the lowest HSS U95$\mid$E95 precision error (0.21~degrees RMS) while matching DiffEyeSyn on RAN (0.29~degrees RMS). However, every synthetic method has substantially lower tail dispersion than the real signals, whose U95$\mid$E95 values are 0.86~degrees RMS for HSS and 1.14~degrees RMS for RAN. This gap suggests that the generators suppress part of the high-variability behavior present in difficult real fixation periods. In practice, this smoothing can be useful for augmentation, gaze-interface simulation, and robust model training, but it may underrepresent extreme tracking noise, calibration error, or atypical oculomotor behavior. Therefore, synthetic sequences should complement rather than replace real recordings when an application must support difficult users or explicitly model worst-case gaze behavior.

The embedding-similarity results provide evidence that EyeMakeYou preserves subject-specific structure beyond visual plausibility. Its cosine similarities range from $0.91$ to $0.95$ across BLG, FXS, HSS, RAN, TEX, VD1, and VD2, exceeding the VAE and SP-EyeGAN baselines and improving on DiffEyeSyn in every reported task. The consistency across fixation, saccade, reading, video-viewing, and gaze-driven interaction tasks suggests that the model does not rely on a single task-specific trajectory pattern. This is an important property for biometric data augmentation, where synthetic samples should retain stable user characteristics across changing visual behaviors. At the same time, EKYT is used to construct the user-identity condition and to compute the identity-preservation loss. Consequently, EKYT cosine similarity should be interpreted as consistency in the EKYT biometric representation, not as a fully independent proof of authentication performance. Future experiments should therefore report the planned EER and Rank-1 identification results under real--real, real--synthetic, and synthetic--real protocols, ideally using an independently trained biometric encoder as an additional evaluator.

The qualitative comparisons are consistent with the quantitative findings. Across HSS, RAN, and TEX, EyeMakeYou better preserves fixation plateaus, the direction and magnitude of major position transitions, and sparse velocity peaks associated with saccades. In contrast, SP-EyeGAN often introduces excessive positional excursions and high-frequency velocity jitter, whereas the VAE frequently attenuates the amplitude and sharpness of transitions. DiffEyeSyn reproduces several major changes but occasionally exhibits channel-specific misalignment and saturated velocity peaks. These observations help explain why EyeMakeYou attains improved median spatial accuracy and high embedding similarity: its generated trajectories better maintain the coupling between horizontal and vertical position changes and their corresponding velocity events. Nevertheless, the qualitative examples are illustrative; they should be interpreted together with the task-wise and user-percentile metrics rather than as evidence of general performance on their own.

The user-state analysis reveals a more nuanced outcome. The synthetic signals retain selected correlations between subjective reports and interpretable oculomotor features, but the correlation structure is weaker, sparser, and less stable across recording rounds than in the real data. HSS and RAN show relatively sparse, round-dependent associations in both conditions, while TEX exhibits the strongest real-data structure and the clearest residual synthetic correlations. These findings indicate partial preservation of task-dependent user-state relationships, especially for reading behavior, but they do not show that the complete real feature--report structure is reproduced. Several factors may contribute to this gap: the user-state conditioning scale is modest, subjective ratings have limited range and floor effects, and the 58-feature Spearman analysis captures marginal feature--report relationships rather than all multivariate or temporal manifestations of state. Accordingly, the correlation analysis should be interpreted as an assessment of how well this specific interpretable feature set preserves reported-state relationships, not as a complete measure of the information represented in the synthetic signal.


Several limitations define the scope of the present findings. First, the evaluation uses 5-s segments from GazeBase recorded with a single eye tracker at 1000~Hz, and the test split contains unseen users but does not establish generalization across devices, laboratories, longer interaction sessions, or naturalistic visual environments. Second, EyeMakeYou is a conditional synthesis method: it receives a low-frequency reference derived from a real gaze recording and an identity embedding extracted from the original velocity signal. Its intended use is therefore conditional augmentation or high-frequency reconstruction of user-specific gaze, rather than unconstrained generation from identity, task, and state labels alone. Third, the current results show meaningful but incomplete preservation of subjective-state relationships, motivating controlled state-manipulation studies and prediction-based state evaluations in addition to correlation analysis. Finally, future work should include repeated random splits, uncertainty estimates, independent biometric evaluators, complete EER and identification experiments, and ablations of the identity, task, state, spectral, drift, and smoothness components.

Overall, EyeMakeYou demonstrates that multi-conditional diffusion can generate high-frequency gaze signals that more closely preserve task-relevant spatial behavior and identity-related embedding structure than the evaluated GAN and VAE baselines. Its strongest use case is as a source of subject-specific synthetic samples for augmenting gaze-based biometric and interactive-system datasets, particularly when real data are limited. The remaining gaps in tail accuracy, extreme variability, event-level feature equivalence, and user-state correlation preservation provide clear targets for future model refinement and evaluation.

\section{Conclusion}

In this work, we presented EyeMakeYou, an identity-, task-, and subjective-state-conditioned diffusion framework for synthesizing subject-specific, high-frequency eye-movement signals. EyeMakeYou generates 5-s, 1000-Hz bivariate gaze-velocity sequences conditioned on a low-frequency reference trajectory, EKYT-derived user embeddings, task embeddings, and subjective-state embeddings, while its training objective combines noise prediction, identity preservation, multi-resolution spectral consistency, drift consistency, and event-weighted local smoothness. Experiments on GazeBase demonstrate that EyeMakeYou achieves strong spatial and biometric fidelity relative to SP-EyeGAN, the VAE baseline, and DiffEyeSyn: it obtains the lowest median spatial errors for the HSS and RAN tasks, achieves the lowest high-percentile fixation jitter for HSS, and yields the highest real--synthetic EKYT embedding similarity across all evaluated tasks. Qualitative results further indicate that the model preserves fixation plateaus, task-dependent position transitions, and saccade-related velocity dynamics. The synthetic user-state analysis identifies task- and round-dependent associations between subjective reports and oculomotor features, with the most distributed associations observed for TEX; however, these relationships remain limited and variable across rounds. Future work will evaluate EyeMakeYou across additional eye trackers, longer recordings, and more naturalistic tasks; report comprehensive biometric authentication and identification performance; and conduct ablations of the identity, task, state, spectral, drift, and smoothness components. These investigations will help establish the role of conditional diffusion in augmenting limited eye-movement datasets and supporting robust gaze-based biometric and interactive systems.


\section*{Privacy and Ethics Statement}
This study uses anonymized, publicly available eye-tracking data to train generative models, posing minimal societal risk. No personal data was used, and we encourage responsible application to avoid potential misuse. Our work supports ethical, privacy-conscious advancements in eye-tracking research.

\bibliographystyle{unsrtnat}
\bibliography{references}  






\end{document}